\documentclass[12pt]{article}
\usepackage{amssymb,amsmath,amsbsy}
\usepackage{epsfig}
\usepackage{array}
\usepackage{float}
\usepackage{amstext}
\usepackage{rotating}
\usepackage{a4}
\usepackage{a4wide}
\usepackage{graphicx}
\usepackage{amsmath} 

\def\be{\begin{equation}}
\def\ee{\end{equation}}
\def\gs{\mathrel{
   \rlap{\raise 0.511ex \hbox{$>$}}{\lower 0.511ex \hbox{$\sim$}}}}
\def\ls{\mathrel{
   \rlap{\raise 0.511ex \hbox{$<$}}{\lower 0.511ex \hbox{$\sim$}}}}

\newcommand{\baz}{\begin{array}{cc}}
\newcommand{\barrr}{\begin{array}{rrr}}
\newcommand{\bad}{\begin{array}{ccc}}
\newcommand{\bav}{\begin{array}{cccc}}
\newcommand{\baf}{\begin{array}{ccccc}}
\newcommand{\bea}{\begin{equation} \begin{array}{c}}
\newcommand{\eea}{ \end{array} \end{equation}}
\newcommand{\D}{\displaystyle}

\newcommand{\meff}{\mbox{$\langle m \rangle$}}

\newcommand{\ba}{\begin{eqnarray}}
\newcommand{\ea}{\end{eqnarray}}

\newcommand{\bs}{\begin{subequations}}
\newcommand{\es}{\end{subequations}}
\newcommand{\no}{\nonumber \\ }
\newcommand{\datm}{\Delta m^2_\mathrm{atm}}
\newcommand{\dsol}{\Delta m^2_\mathrm{sol}}

\newcommand{\gsim}{\raise0.3ex\hbox{$\;>$\kern-0.75em\raise-1.1ex\hbox{
   $\sim\;$}}} 
\newcommand{\lsim}{\raise0.3ex\hbox{$\;<$\kern-0.75em\raise-1.1ex\hbox{
   $\sim\;$}}}

\begin{document}

\title{
\hfill{\small CFTP/13-016}\\[1cm]
\Large \bf Reproducing lepton mixing in a texture zero model }
\author{
Lu\'\i s Lavoura,$^a$\thanks{email: 
\tt balio@cftp.ist.utl.pt}~
Werner Rodejohann,$^b$\thanks{email: 
\tt werner.rodejohann@mpi-hd.mpg.de}~
and Atsushi Watanabe$^b$\thanks{email: 
\tt atsushi.watanabe@mpi-hd.mpg.de}
\\\\
{\normalsize \it $^a$University of Lisbon,
Instituto Superior T\'ecnico,
CFTP,}\\
{\normalsize \it 1049-001 Lisboa, Portugal}\\ \\
{\normalsize \it $^b$Max--Planck--Institut f\"ur Kernphysik,}\\
{\normalsize \it Saupfercheckweg 1, D--69117 Heidelberg, Germany}}
\date{}
\maketitle
\thispagestyle{empty}
\vspace{0.8cm}
\begin{abstract}
\noindent  
We note that the emerging features of lepton mixing can be reproduced if,
with inverted neutrino mass ordering,
both the smallest neutrino mass
and the $\tau\tau$ element of the neutrino mass matrix
vanish.
Then,
the atmospheric neutrino mixing angle is less than maximal
and the Dirac phase $\delta$ is close to $\pi$.
We derive the correlations among the mixing parameters and show that
there is a large cancellation
in the effective mass responsible for neutrinoless $\beta\beta$ decay.
Three simple seesaw models leading to our scenario are provided.

\end{abstract}

\newpage

The increasingly precise determination of the lepton mixing parameters
makes a theoretical interpretation of the data a necessity.
In this Letter we propose a form of the neutrino mass matrix
that accommodates,
and in fact correlates, 
two emerging structures~\cite{Fogli:2012ua} of lepton mixing: 
(i) the value of the atmospheric mixing angle $\theta_{23}$
is significantly less than maximal;
(ii) the $CP$-violating Dirac phase $\delta$ is close to $\pi$. 
Table~\ref{tab:osc} shows the mixing parameters
obtained from a global fit
to the neutrino oscillation data.\footnote{Other fits~\cite{tortola,schwetz}
give somewhat different results for $\theta_{23}$ and for $\delta$.}
\begin{table}[ht]
\centering
\begin{tabular}{|l|c|c|c|}
\hline
Parameter & best-fit & $1\sigma$ range  & $3\sigma$ range
\\[.15cm] \hline \hline
$\dsol \, [10^{-5}~\mathrm{eV}^2] $  & 7.54 & 7.32 -- 7.80 & 6.99 -- 8.18
\\[.12cm] \hline
$\datm \, [10^{-3}~\mathrm{eV}^2] $ & 2.42 & 2.31 -- 2.49  & 2.17 -- 2.61
\\ [.12cm] \hline
$\sin^2 \theta_{12}$  & 0.307 & 0.291 -- 0.325  & 0.259 -- 0.359
\\ [.12cm] \hline
$\sin^2 \theta_{13}$ & 0.0244 & 0.0219 -- 0.0267  & 0.0171 -- 0.0315
\\ [.12cm] \hline
$\sin^2 \theta_{23}$ & 0.392 & 0.370 -- 0.431 & 0.335 -- 0.663
\\ [.12cm] \hline
$\delta$ & $1.09\pi$ & ($0.83$ -- $1.47$)$\pi$  & 0 -- $2 \pi$
\\ [.12cm] \hline
\end{tabular}
\caption{\label{tab:osc}Values of the lepton mixing parameters
in the case of an inverted neutrino mass ordering.
This table was taken in abridged form from Ref.~\cite{Fogli:2012ua}.}
\end{table}

While features~(i) and~(ii) are not fully established yet,
they seem interesting and worth investigating.
The proposal that we make is that,
in the basis of a diagonal charged-lepton mass matrix,\footnote{Actually,
it is enough that the diagonalization of the charged-lepton mass matrix
consists only of a rotation in the $e$--$\mu$ plane.}
the following features hold:
\be \label{eq:main}
M_{\tau\tau} = m_3 = 0 \, , 
\ee
where $M$ is the neutrino mass matrix and
$m_3$ is the mass of the lightest neutrino
when the neutrino mass hierarchy is inverted.\footnote{It could already
be gathered from Ref.~\cite{Liao:2013rca} that
Eq.~(\ref{eq:main}) is in agreement with the data.}
We shall show that this leads to the correlation
\be
\label{ibupe}
\tan{\theta_{23}} \simeq
- 2 \sin{\theta_{13}} \cos{\delta} \tan{2 \theta_{12}}\, , 
\ee
and therefore implies the desired features that
$\cos{\delta}$ is negative and large
while $\theta_{23}$ is small,
hence significantly less than maximal.\footnote{The so-called TM$_1$ scenario, 
in which the first column of the PMNS matrix satisfies 
$\left( \left| U_{e1} \right|, \left| U_{\mu 1} \right|,
\left| U_{\tau 1} \right| \right)
= \left( \sqrt{\frac 2 3}, \sqrt{\frac 1 6}, \sqrt{\frac 1 6} \right)$, 
displays~\cite{Albright:2008rp} a similar correlation between $\theta_{23}$
and $\delta$.}
Moreover,
our scheme makes predictions:
(i) there is an inverted neutrino mass hierarchy with vanishing smallest mass;
(ii) the sole Majorana phase has a value that leads
to a large cancellation in the effective mass on which the lifetime
of neutrinoless $\beta \beta$ decay depends,
hence that lifetime must lie at the upper end of its allowed range. 

The implications of single texture zeros,
with and without vanishing smallest neutrino mass,
have often been studied~\cite{Frampton:2002yf,
Merle:2006du,
Guo:2006qa,
Lashin:2011dn,
Ludl:2011vv, Rodejohann:2012jz,Grimus:2012ii}.
We stress in this Letter how well the consequences of Eq.~(\ref{eq:main})
fit the current data. 
We also point out that Eq.~(\ref{eq:main}) can be arranged in models;
we illustrate this through three seesaw models
furnished with discrete symmetries.

We start with the phenomenology of Eq.~(\ref{eq:main}).
Defining---as in Ref.~\cite{Fogli:2012ua},
from which we have taken the values of the mixing
parameters---$\datm =
\left| m_3^2 - \left. \left( m_1^2 + m_2^2 \right) \right/ 2 \right|$,
one has,
from $m_3 = 0$,
that
\bs
\ba
m_1^2 &=& \datm \left( 1 - \frac{r}{2} \right),
\\
m_2^2 &=& \datm \left( 1 + \frac{r}{2} \right),
\ea
\es
where $r \equiv \dsol / \datm \approx 0.031$
and $\dsol = m_2^2 - m_1^2$. 
Since the neutrino mass matrix,
in the charged-lepton mass basis,
is $M = U^\ast\, \mathrm{diag} \left( m_1, m_2, m_3 \right) U^\dagger$,
where $U$ is the PMNS matrix,
one has 
\be
\label{eme33}
M_{\tau\tau} = 0 \,\Leftrightarrow\,
U_{\tau 1}^2 \sqrt{1 - \frac{r}{2}}
= - U_{\tau 2}^2 \sqrt{1 + \frac{r}{2}}\, .
\ee
In the standard parametrization of $U$,
%
\bs \label{ivuoq}
\ba
U_{\tau 1} &=& s_{12} s_{23} - c_{12} c_{23} s_{13} e^{i \delta} \, ,
\\*[0.2mm]
U_{\tau 2} &=& \left( - c_{12} s_{23} - s_{12} c_{23} s_{13} e^{i \delta} \right) 
e^{i \rho / 2} \, ,
\ea
\es
where $s_j = \sin{\theta_j}$ and $c_j = \cos{\theta_j}$
for $j = 12, 13, 23$.
To get rid of the Majorana phase $\rho$ one
takes the moduli of both sides of Eq.~(\ref{eme33}):
\bea \D 
\left( s_{12}^2 s_{23}^2 + c_{12}^2 c_{23}^2 s_{13}^2
- 2 s_{12} c_{12} s_{23} c_{23} s_{13} \cos{\delta} \right)
\sqrt{1 - \frac{r}{2}} \\ \D 
 = \left( c_{12}^2 s_{23}^2 + s_{12}^2 c_{23}^2 s_{13}^2
+ 2 s_{12} c_{12} s_{23} c_{23} s_{13} \cos{\delta} \right)
\sqrt{1 + \frac{r}{2}}\,.
\label{eme332}
\eea
Defining 
\be
\varepsilon \equiv \frac
{\sqrt{1 - r/2} - \sqrt{1 + r/2}}
{\sqrt{1 - r/2} + \sqrt{1 + r/2}}
= - \frac{r}{4} - \frac{r^3}{64} - 
\cdots, 
\ee
equation~(\ref{eme332}) may be rewritten as 
\be
\left( s_{12}^2 - c_{12}^2 \right) \left( s_{23}^2 - c_{23}^2 s_{13}^2 \right)
- 4 s_{12} c_{12} s_{23} c_{23} s_{13} \cos{\delta}
+ \varepsilon \left( s_{23}^2 + c_{23}^2 s_{13}^2 \right) = 0\,.
\label{eme333}
\ee
Since $c_{12}^2 - s_{12}^2 \approx 0.4$ while $\varepsilon \approx 0.008$,
the third term in the left-hand side of Eq.~(\ref{eme333})
may be neglected relative to the first one
and one ends up with\footnote{The exact version of Eq.~(\ref{cd1}) is
\be
\cos{\delta} = - \frac{\left( c_{12}^2 - s_{12}^2 \right) s_{23}^2
- s_{13}^2 \left( c_{12}^2 - s_{12}^2 \right) c_{23}^2
+ \varepsilon \left( s_{23}^2 + c_{23}^2 s_{13}^2 \right)}
{4 s_{12} c_{12} s_{23} c_{23} s_{13}} \, ,
\ee
The first term of the numerator is of order 0.15
while the second and third terms are of order 0.006 and 0.004,
respectively.}
\be
\label{cd1}
\cos{\delta} \approx - \frac{\tan{\theta_{23}}}
{2  \sin \theta_{13} \tan{ 2 \theta_{12} }}\,,
\ee
and we have demonstrated Eq.~(\ref{ibupe}).
Equation~(\ref{cd1})
indicates that $\cos{\delta}$ is negative.
Moreover,
since $\sin \theta_{13} \approx 0.16$ is small,
$\left| \cos{\delta} \right|$ should be large.
In order for it not to exceed 1,
$\theta_{23}$ must be as small as possible,
hence in the first octant,
while both $\theta_{13}$ and $\theta_{12}$ should be 
largish. To be precise, 
\begin{eqnarray}
\sin{\theta_{13}} \tan{2\theta_{12}}
> \frac{\left. \tan{\theta_{23}} 
\right|_{\rm min \left( 3\sigma \right)}}{2} = 0.355
\label{relation2}
\end{eqnarray}
must hold.

The arguments of the left- and right-hand sides of Eq.~(\ref{eme33})
should also coincide.
Neglecting the small terms containing $s_{13}$ in Eqs.~(\ref{ivuoq}),
this results in the condition 
\be \label{eq:rho}
\rho \approx \pi \,. 
\ee
A consequence of Eq.~(\ref{eq:rho})
occurs in the effective mass $|M_{ee}| \equiv \meff$ that governs the rate
of neutrinoless $\beta \beta$ decay~\cite{Rodejohann:2012xd}.
For vanishing $m_3$,
and for $\rho \approx \pi$,
this is given by 
\be
\meff \approx c_{13}^2 \sqrt{\datm}
\sqrt{1 - \sin^2{2 \theta_{12}} \, \sin^2{\rho/2}}
\stackrel{\rho \approx \pi}{\approx}
c_{13}^2 \sqrt{\datm} \cos{2 \theta_{12}} \, . 
\ee 
This gives $\meff \approx 0.018\,{\rm eV}$
for the best-fit values in Table~\ref{tab:osc}.
The effective mass therefore lies at the lower end of the range 
$0.013\,{\rm eV} < \meff < 0.050\,{\rm eV}$
generally allowed for the inverted hierarchy. 
Current limits on $\meff$ are around 0.3 eV,
while future experiments are aiming
at entering the regime of the inverted hierarchy
by improving current lifetime limits on neutrinoless $\beta \beta$ decay 
by an order of magnitude.  

Regarding other neutrino mass observables,
KATRIN~\cite{Drexlin:2013lha,Parno} will not be able to see a signal
if our scheme holds,
whereas in cosmology~\cite{Abazajian:2011dt}
there could be detection in sophisticated future surveys,
since the sum of the light-neutrino masses is 
\be
m_1 + m_2 \approx 2 \sqrt{\datm} \approx 0.1\, \mathrm{eV} \, . 
\ee

\begin{figure}[h]
\begin{center}
\includegraphics[width=6cm,height=6cm]{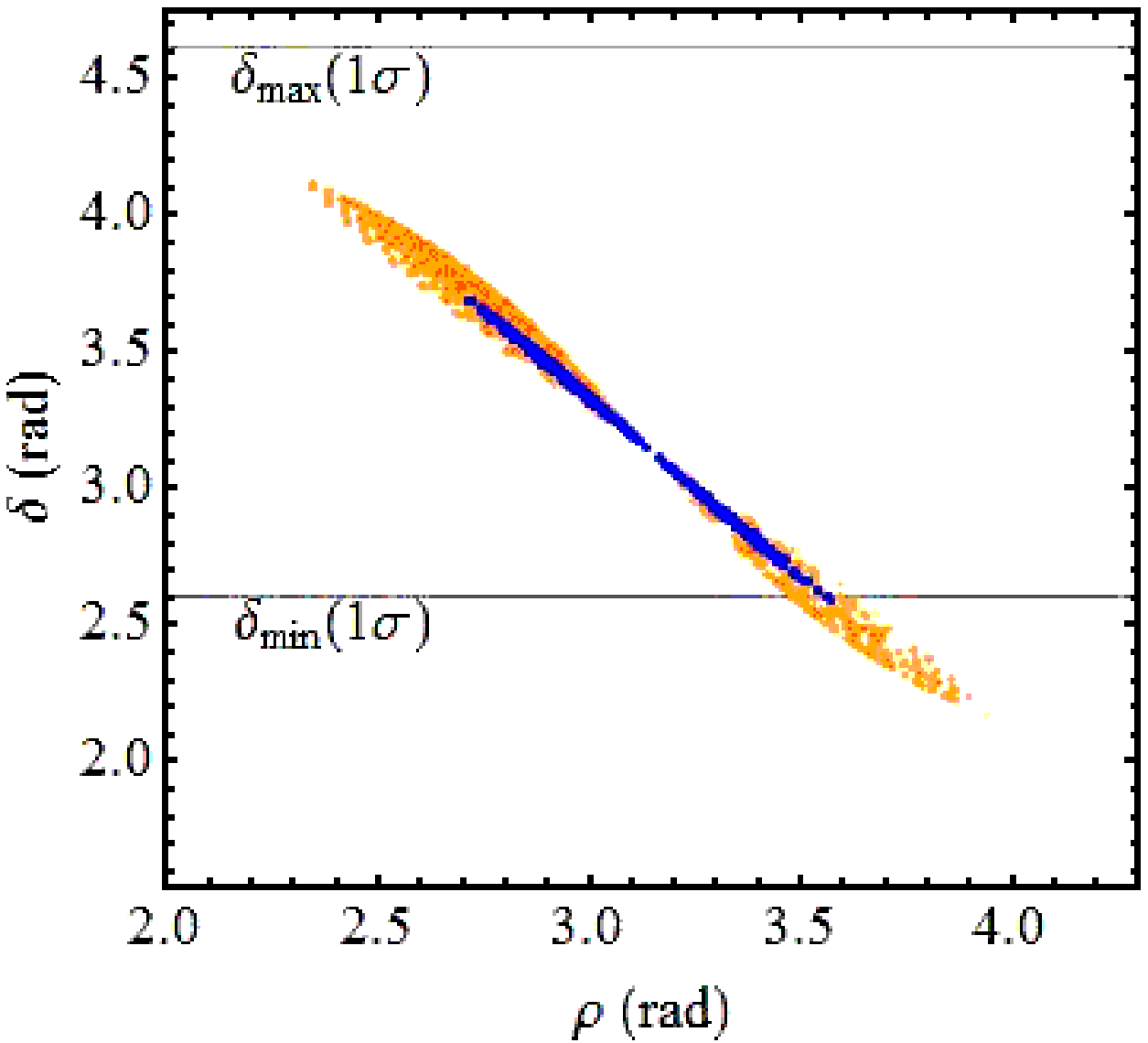}
\includegraphics[width=6cm,height=6cm]{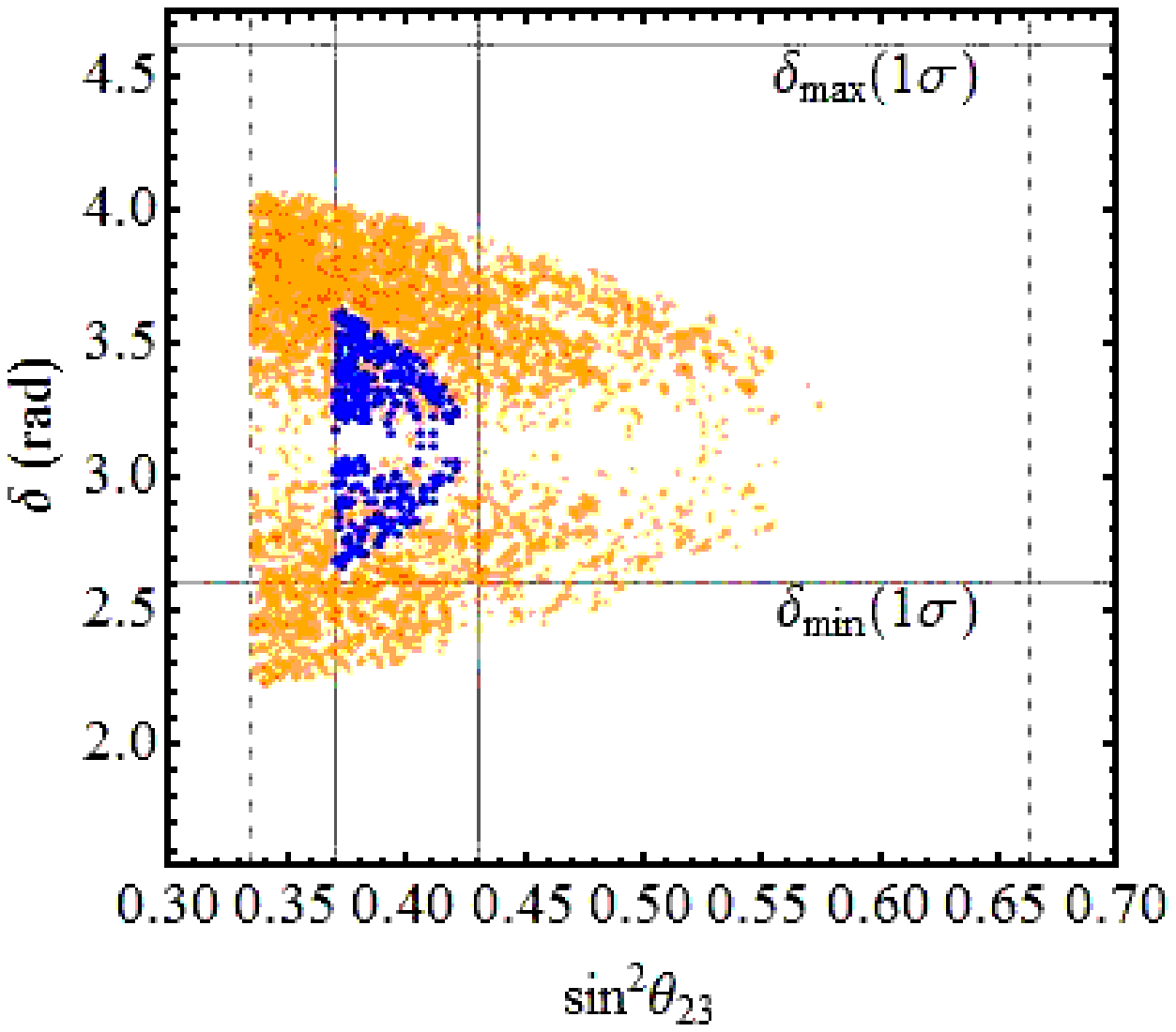}

\vspace{7mm}

\includegraphics[width=6cm,height=6cm]{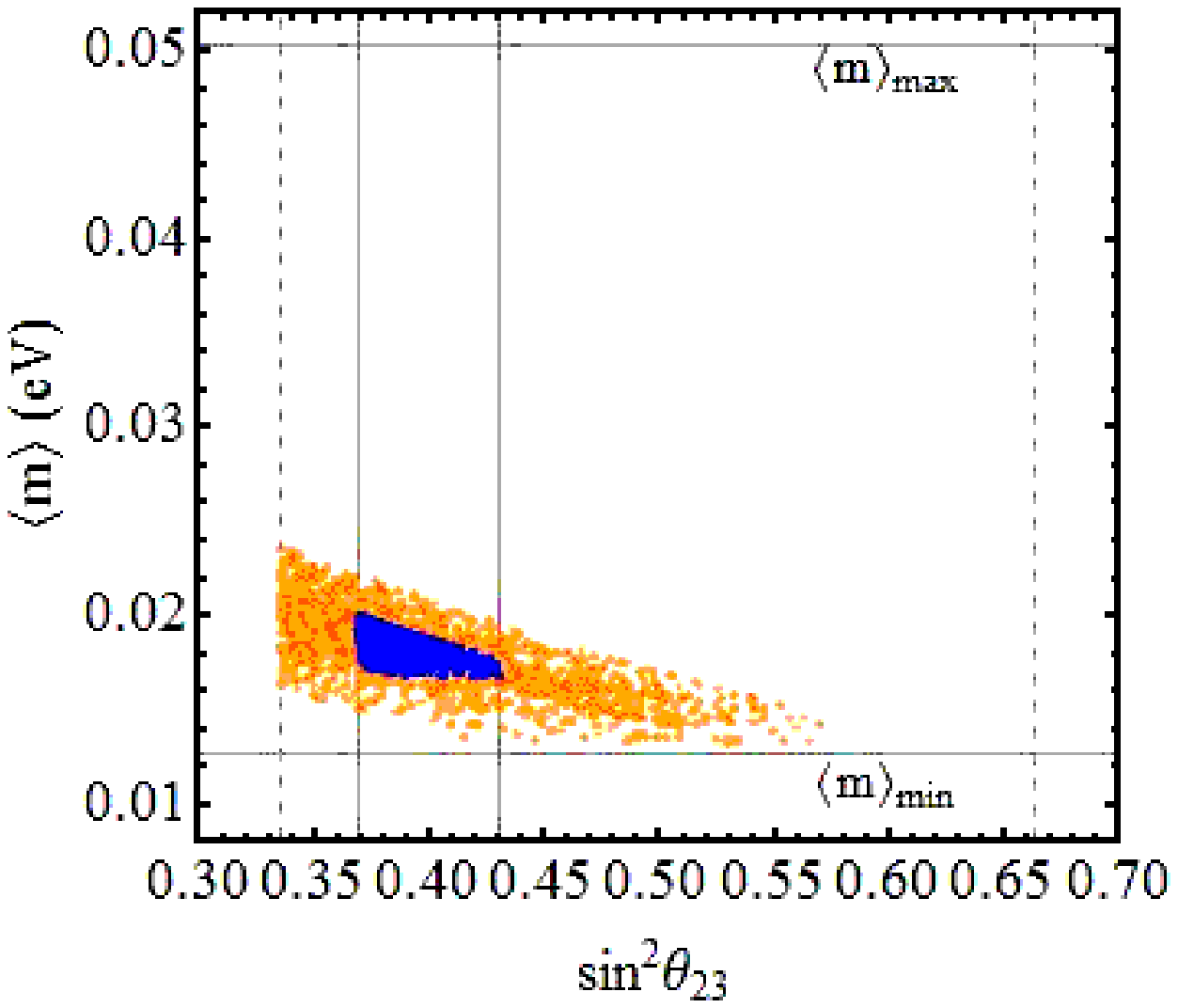}
\includegraphics[width=6cm,height=6cm]{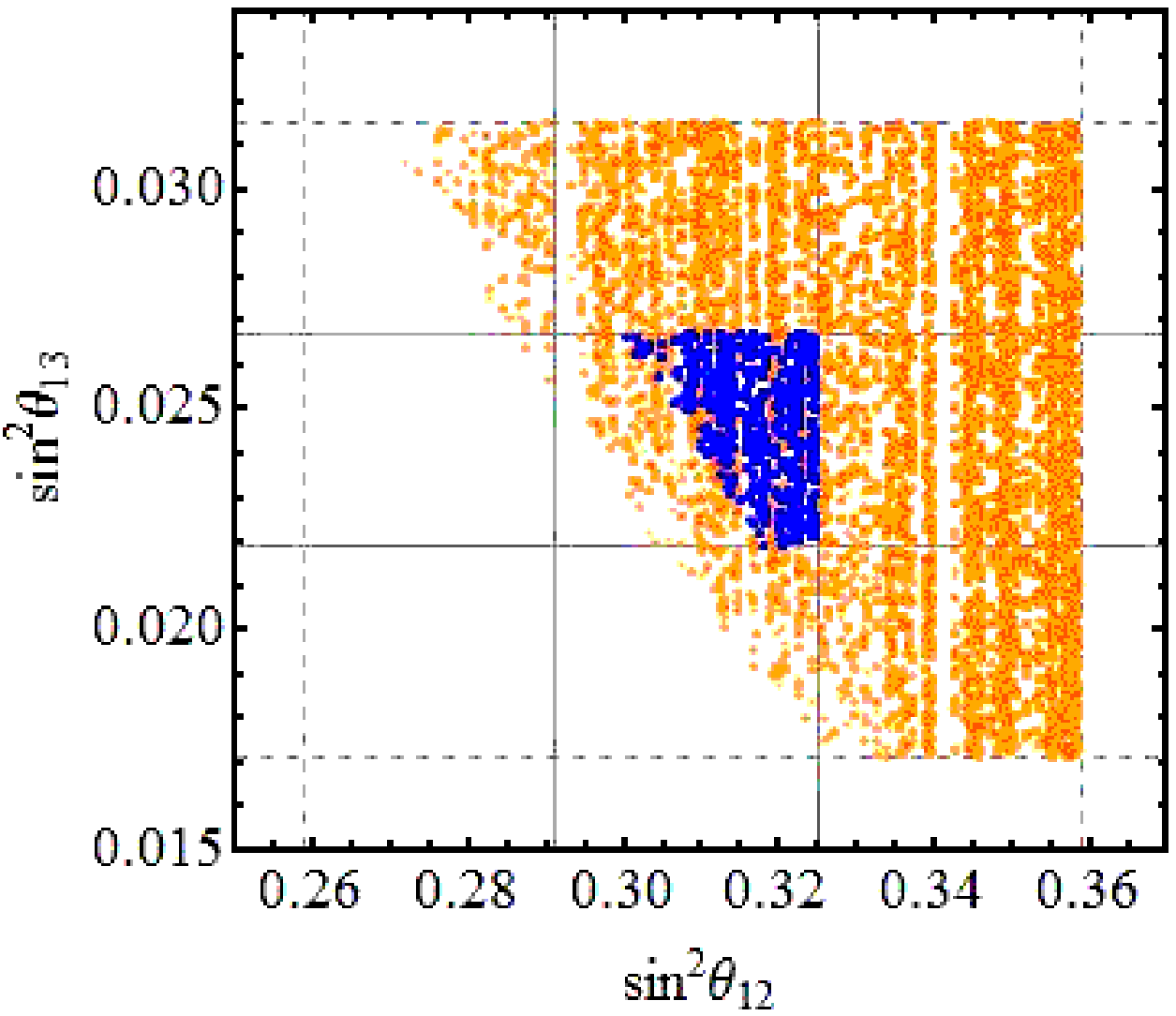}
\end{center}
\caption{Scatter plots displaying
the correlations among some neutrino mass and mixing observables
following from Eq.~(\ref{eq:main}).
The horizontal and vertical full lines
display the $1\sigma$ bounds of Ref.~\cite{Fogli:2012ua}
on the corresponding observables;
the dashed lines display the $3\sigma$ bounds.
The blue (dark) points were found
by allowing all the observables to lie within
their corresponding $1\sigma$ ranges of Table~\ref{tab:osc};
the yellow (light) points pertain to the $3\sigma$ ranges.
$\langle m \rangle_{\rm min}$ and $\langle m \rangle_{\rm max}$ are the minimal and 
maximal values of the effective mass in the inverted hierarchy with vanishing 
neutrino mass.}
\label{fig:pheno}
\end{figure}

We display the phenomenology of our scenario in Fig.~\ref{fig:pheno},
which confirms our analytical expressions.
Again,
we stress the interesting correlation 
between the $CP$-violating Dirac phase $\delta$
and the atmospheric neutrino mixing angle $\theta_{23}$.

\vspace*{3mm}

Next we turn to model realizations of the scenario under study.
We shall present three such models.
We use the type-I seesaw mechanism
with only two right-handed neutrinos $\nu_{R1,2}$;
this immediately ensures the existence of one massless neutrino.
In the first model we need two Higgs doublets $\phi_{1,2}$
and a complex scalar gauge singlet $S$.
Let there be a $\mathbb{Z}_4$ symmetry under which
the fields---including the left-handed lepton doublets
$D_{L\alpha}$ ($\alpha = e, \mu, \tau$)---transform as
\bs
\label{uvyur}
\ba
& D_{Le} \to - i D_{Le}\,, \quad
D_{L\mu} \to - i D_{L\mu}\,, \quad
e_R \to - e_R\, , \quad
\mu_R \to - \mu_R\, , &
\label{uvirt}
\\*[1mm]
& \nu_{R1} \to - i \nu_{R1}\,, \quad
\phi_1 \to i \phi_1\,, \quad
S \to - i S\,. &
\ea
\es
The Yukawa couplings $\bar D_{Le} \phi_1 e_R$
and $\bar D_{L\mu} \phi_1 \mu_R$
give mass to the electron and the muon
while $\bar D_{L\tau} \phi_2 \tau_R$ gives mass
to the $\tau$.\footnote{Notice that,
without loss of generality,
we may assume the charged-lepton mass matrix already to be diagonal,
since its diagonalization only amounts to a redefinition
of $\left( D_{Le}, D_{L\mu} \right)$
and of $\left( e_R, \mu_R \right)$ in Eq.~(\ref{uvirt}).}
The
Yukawa couplings to the right-handed neutrinos are
\ba
\mathcal{L}_\mathrm{Yukawa} &=&
- y_6 \, \bar \nu_{R1} C \bar \nu_{R2}^T \,S
- \left( y_1 \bar D_{Le} + y_2 \bar D_{L\mu} \right) \nu_{R1}\, \tilde \phi_2
\no & &
- \left( y_4 \bar D_{Le} + y_5 \bar D_{L\mu} \right) \nu_{R2}\, \tilde \phi_1
- y_3 \bar D_{L\tau} \,\nu_{R2}\, \tilde \phi_2
+ \mathrm{H.c.},
\label{mvjut}
\ea
where $y_{1\mbox{--}6}$ are coefficients,
$C$ is the charge-conjugation matrix in Dirac space,
and $\tilde \phi_k = i \sigma_2 \phi_k^\ast$ for $k = 1, 2$.
There is also a bare Majorana mass term
\be
\mathcal{L}_\mathrm{Majorana} = - \frac{m}{2}\,
\bar \nu_{R2} \,C \bar \nu_{R2}^T
+ \mathrm{H.c.}
\ee
Therefore,
the neutrino mass matrices,
in the notation
\be
\mathcal{L}_{\nu\, \mathrm{mass}} = - \bar \nu_R \,M_D \,\nu_L
- \frac{1}{2}\, \bar \nu_{R} \,M_R \,C \bar \nu_R^T + \mathrm{H.c.},
\ee
are
\be
M_D = \left( \begin{array}{ccc}
y_1^\ast v_2 & y_2^\ast v_2 & 0 \\ y_4^\ast v_1 & y_5^\ast v_1 & y_3^\ast v_2
\end{array} \right),
\quad
M_R = \left( \begin{array}{cc}
0 & y_6 s \\ y_6 s & m
\end{array} \right),
\ee
where $s = \left\langle 0 \left| S \right| 0 \right\rangle$
and $v_k = \left\langle 0 \left| \phi_k^0 \right| 0 \right\rangle$.
The effective light-neutrino mass matrix is
\ba
M &=& - M_D^T \,M_R^{-1} \,M_D
\no &=& \frac{- 1}{y_6^2 s^2}
\left( \begin{array}{cc}
y_1^\ast v_2 & y_4^\ast v_1 \\ y_2^\ast v_2 & y_5^\ast v_1 \\ 0 & y_3^\ast v_2
\end{array} \right)
\left( \begin{array}{cc}
- m & y_6 s \\ y_6 s & 0
\end{array} \right)
\left( \begin{array}{ccc}
y_1^\ast v_2 & y_2^\ast v_2 & 0 \\ y_4^\ast v_1 & y_5^\ast v_1 & y_3^\ast v_2
\end{array} \right),
\ea
which evidently has $M_{\tau\tau} = 0$.

\vspace*{3mm}

A simpler model,
which dispenses with the singlet $S$,
is the following.
Let there be a symmetry under which
\bs
\ba
&
D_{Le} \to \sigma D_{Le}\, \, \quad
D_{L\mu} \to \sigma D_{L\mu}\, , \quad
D_{L\tau} \to \sigma^{-1} D_{L\tau}\,,
&
\\*[1mm]
&
e_R \to \sigma^3 e_R\, , \quad
\mu_R \to \sigma^3 \mu_R\, , \quad
\tau_R \to \sigma^{-1} \tau_R\, , 
&
\\*[1mm]
&
\nu_{R1} \to \sigma \nu_{R1}\,, \quad
\nu_{R2} \to \sigma^{-1} \nu_{R2}\,, \quad
\phi_1 \to \sigma^{-2} \phi_1\,.
&
\ea
\es
where $\left| \sigma \right| = 1$ and $\sigma^4 \neq 1$.
Then the Yukawa couplings to the right-handed neutrinos are
\ba
\mathcal{L}_\mathrm{\nu\, Yukawa} &=&
- \left( y_1 \bar D_{Le} + y_2 \bar D_{L\mu} \right) \nu_{R1} \,\tilde \phi_2
\no & &
- \left( y_4 \bar D_{Le} + y_5 \bar D_{L\mu} \right) \nu_{R2} \,\tilde \phi_1
- y_3 \bar D_{L\tau} \,\nu_{R2} \,\tilde \phi_2
+ \mathrm{H.c.},
\ea
where $y_{1\mbox{--}5}$ are coefficients.
There is a bare Majorana mass term
\be
\mathcal{L}_\mathrm{Majorana} = - m \,\bar \nu_{R1} \,C \bar \nu_{R2}^T
+ \mathrm{H.c.}
\ee
Defining 
\be
a \equiv y_1^\ast v_2\,, 
b \equiv y_2^\ast v_2\,, 
c \equiv y_3^\ast v_2\,, 
d \equiv y_4^\ast v_1\,, 
f \equiv y_5^\ast v_1\,,
\ee
gives for the mass matrix
\be
M =
- \frac{1}{m}
\left( \begin{array}{cc} a & d \\ b & f \\ 0 & c \end{array} \right)
\left( \begin{array}{cc} 0 & 1 \\ 1 & 0 \end{array} \right)
\left( \begin{array}{ccc} a & b & 0 \\ d & f & c \end{array} \right)
= - \frac{1}{m} \left( \begin{array}{ccc}
2 a d & a f + b d & a c \\
a f + b d & 2 b f & b c \\
a c & b c & 0
\end{array} \right).
\ee
This $M$ contains five physical parameters, 
\be
\left| \frac{a d}{m} \right|, \quad
\left| \frac{b}{a} \right|, \quad
\left| \frac{f}{d} \right|, \quad
\left| \frac{c}{d} \right|, \quad
\arg{\left( a f b^\ast d^\ast \right)}\,.
\ee
These parameters can be computed from the matrix elements of $M$,
and thus from the neutrino masses and mixings.
Notice for instance that
\be
\left| \frac{y_1}{y_2} \right| = \left| \frac{M_{e\tau}}{M_{\mu\tau}} \right|,
\quad
\left| \frac{y_4}{y_5} \right| =
\left| \frac{M_{ee} M_{\mu\tau}}{M_{\mu\mu} M_{e\tau}} \right|.
\ee

\vspace*{3mm}

Another example of a model leading to $M_{\tau\tau}=0$ uses 
$S_3$ and $\mathbb{Z}_3$ discrete groups as flavour symmetries.
The multiplication rules of $S_3$
can be found for instance in Ref.~\cite{Rodejohann:2012jz}. 
The particle content and group assignments are
given in Table~\ref{table:model3}.
\begin{table}[ht]
\centering
\begin{tabular}{|l||c|c||c||c|c|c||c|c|}
\hline
  &
$\left( D_{Le}, D_{L\mu} \right)$ & $D_{L\tau}$ &
$ \left( \nu_{R1}, \nu_{R2} \right)$ &
$e_R$ & $\mu_R$ & $\tau_R$ &
$\left( \xi_1, \xi_2 \right)$ & $\chi$
\\
\hline \hline
$S_3$ &
$\mathbf{2}$ & $\mathbf{1}'$ &
$\mathbf{2}$ &
$\mathbf{1}$ & $\mathbf{1}$ & $\mathbf{1}'$ &
$\mathbf{2}$ & $\mathbf{1}$
\\
\hline
$\mathbb{Z}_3$ &
$1$ & $1$ &
$\omega$ &
$\omega^2$ & $\omega$ & $1$ & 
$\omega$ & $\omega$
\\
\hline
\end{tabular}
\caption{
Field content of our third model.
The standard-model Higgs doublet $\phi$,
which is invariant under both $S_3$ and $\mathbb{Z}_3$,
has been omitted.
The fields $\xi_{1,2}$ and $\chi$ are `familons',
\textit{i.e.}\ auxiliary scalar gauge singlets.
We use $\omega \equiv \exp{\left( i 2 \pi / 3 \right)}$.
}
\label{table:model3}
\end{table}

At leading order in the inverse power of the cutoff scale $\Lambda$,
the neutrino Yukawa couplings and the Majorana masses are given by
\bs
\ba
\mathcal{L}_\nu &=&
- \frac{g_1}{2} \left(
\bar \nu_{R1} \,C \bar \nu_{R1}^T\, \xi_1^\ast
+
\bar \nu_{R2} \,C \bar \nu_{R2}^T\, \xi_2^\ast
\right)
- g_2\, \bar \nu_{R1} \,C \bar \nu_{R2}^T\, \chi^\ast
\\ & &
- h_1 \left(
\bar D_{Le} \,\nu_{R1} + \bar D_{L\mu}\,\nu_{R2} \right)
\chi^*\, \frac{\tilde \phi}{\Lambda}
\\ & &
- h_2 \left( \bar D_{Le}\,\nu_{R2} \,\xi_1^* + \bar D_{L\mu}\,\nu_{R1}\,\xi_2^* \right)
\frac{\tilde \phi}{\Lambda}
\\ & &
- h_3\, \bar D_{L\tau}
\left( \nu_{R1} \,\xi_1^* - \nu_{R2}\,\xi_2^* \right)
\frac{\tilde \phi}{\Lambda}
+ {\rm H.c.},
\ea
\es
where $g_{1,2}$ and $h_{1,2,3}$ are dimensionless constants.
After $\xi_{1,2}$ and $\chi$ obtain vacuum expectation values according
to the alignment~\cite{Haba:2005ds}
\be
\begin{pmatrix} \xi_1 \\ \xi_2 \end{pmatrix}
\to 
\begin{pmatrix} \langle \xi_1 \rangle \\ 0 \end{pmatrix},
\quad
\chi \to \langle \chi \rangle,
\label{vac1}
\ee
and after electroweak symmetry breaking with $v = \langle \tilde \phi \rangle$, 
the mass matrices at low energy take the forms
\be
M_D = \begin{pmatrix}
a & b \\
0 & a \\
c & 0
\end{pmatrix},
\quad
M_R = \begin{pmatrix}
M_A & M_B \\
M_B & 0 \\
\end{pmatrix},
\label{mDMR1}
\ee
where $a = h_1 \langle \chi \rangle^* v/\Lambda$, 
$b = h_2 \langle \xi_1 \rangle^* v/\Lambda$, $c = h_3 \langle \xi_1 \rangle^* v/\Lambda$,
$M_A = g_1 \langle \xi_1 \rangle^\ast$
and $M_B = g_2 \langle \chi \rangle^\ast$.
Performing the seesaw diagonalization,
one finds the effective Majorana mass matrix 
\be
M = 
\frac{1}{M_B}
\begin{pmatrix}
2ab & a^2 & bc \\
a^2& 0 & ac \\
bc & ac & 0
\end{pmatrix}
- \frac{M_A}{M_B^2}
\begin{pmatrix}
b^2 & ab &0 \\
ab & a^2 & 0 \\
0 & 0 & 0 \\
\end{pmatrix}.
\label{mnu1}
\ee
The charged-lepton mass sector is written through
a combination of higher-dimensional operators
and a renormalizable operator.
At leading order in $1 / \Lambda$,
it is given by
\bs
\ba
\mathcal{L}_l &=& 
- h_\tau \bar D_{L\tau} \,\tau_R \,\phi
\\ & &
- h_e \left(
\bar D_{Le} \,\xi_1
+
\bar D_{L\mu} \,\xi_2
\right) e_R\, \frac{\phi}{\Lambda}
\\ & &
- h_\mu \left(
\bar D_{Le} \,\xi_2^\ast + \bar D_{L\mu} \,\xi_1^\ast
\right)
\mu_R\, \frac{\phi}{\Lambda}
+ {\rm H.c.},
\ea
\es
where $h_{e,\mu,\tau}$ are dimensionless constants.
In the vacuum configuration of Eq.~(\ref{vac1}),
the charged-lepton mass matrix is diagonal:
\be
M_l = \mathrm{diag} \left(
\frac{h_e}{\Lambda}\, \langle \xi_1 \rangle,\,
\frac{h_\mu}{\Lambda}\, \langle \xi_1 \rangle^\ast,\,
h_\tau 
\right) v^\ast.
\label{ml1}
\ee
The hierarchy between $m_{e,\mu}$ and $m_\tau$ is naturally explained
by the suppression by the cutoff scale.

A dark matter candidate is easily accommodated in this model.
For example, we can replace the $S_3$ charges of $D_{L\tau}$ and $\tau_R$
with $\mathbf{1}$ and introduce a third right-handed neutrino, $\nu_{R3}$,
which is $\mathbf{1}'$ of $S_3$ and $\omega$ of $\mathbb{Z}_3$. 
Then, 
$\nu_{R3}$ has no Yukawa couplings,
so that it is stable. 
In the early Universe,
$\nu_{R3}$ can communicate with the thermal plasma
via the $s$-channel exchange of the scalar fields,
since it has a $\nu_{R3} \nu_{R3} \chi$ vertex
and the real part of $\chi$ mixes with the usual Higgs field.
In this setup,
it is known~\cite{LopezHonorez:2012kv}
that the observed relic density is easily obtained
without contradicting the Higgs properties observed at LHC
and the direct detection bounds.

\vspace*{3mm}
\newpage
In summary,
a simple texture-zero scenario
can accommodate the values of the $CP$ phase $\delta$ around $\pi$
and of the atmospheric mixing angles $\theta_{23}$
sizably less than $\pi / 4$.
The framework makes predictions
in the form of an inverted hierarchy with a massless neutrino
and a strong cancellation
in neutrinoless $\beta \beta$ decay.  
Simple models are possible to realize the scenario.

\vspace{0.3cm}
\begin{center}
{\bf Acknowledgments}
\end{center}
The work of LL is supported through
the Marie Curie Initial Training Network
``UNILHC''
PITN-GA-2009-237920
and also through the projects PEst-OE-FIS-UI0777-2011,
PTDC-FIS-098188-2008,
and CERN-FP-123580-2011 of the Portuguese FCT.
The work of WR was supported by the Max Planck Society
through the Strategic Innovation Fund in the project MANITOP.

\end{document}